\documentclass[runningheads]{llncs}
\usepackage{graphicx}
\usepackage{enumitem}
\usepackage{array}
\usepackage{multirow}
\usepackage{booktabs}
\usepackage[normalem]{ulem}
\graphicspath{{figs/}}
\usepackage[misc]{ifsym}
\usepackage{amsmath,amssymb,amsfonts,bm}
\useunder{\uline}{\ul}{}
\begin{document}
\title{Improving Sequential Recommendation with Attribute-augmented Graph Neural Networks}
\titlerunning{ }
%
\author{Xinzhou Dong\inst{1,2} \and
	Beihong Jin\inst{1,2 \textrm{(\Letter)}} \and
	Wei Zhuo\inst{3}\and
	Beibei Li\inst{1,2} \and
	Taofeng Xue \inst{1,2}}
%
\authorrunning{ }
%
\institute{State Key Laboratory of Computer Science, Institute of Software, Chinese Academy of Sciences, Beijing, China \\ \email{Beihong@iscas.ac.cn} \and
	University of Chinese Academy of Sciences, Beijing, China \and
	MX Media Co., Ltd, Singapore}
%
\maketitle              
\begin{abstract}
Many practical recommender systems provide item recommendation for different users only via mining user-item interactions but totally ignoring the rich attribute information of items that users interact with. In this paper, we propose an attribute-augmented graph neural network model named Murzim. Murzim takes as input the graphs constructed from the user-item interaction sequences and corresponding item attribute sequences. By combining the GNNs with node aggregation and an attention network, Murzim can capture user preference patterns, generate embeddings for user-item interaction sequences, and then generate recommendations through next-item prediction. We conduct extensive experiments on multiple datasets. Experimental results show that Murzim outperforms several state-of-the-art methods in terms of recall and MRR, which illustrates that Murzim can make use of item attribute information to produce better recommendations. At present, Murzim has been deployed in MX Player, one of India's largest streaming platforms, and is recommending videos for tens of thousands of users.

\end{abstract}
\section{Introduction}
Sequential recommendation is to predict the next item that a user is most likely to interact with according to the user-item interaction sequence over a period of time in the past and then recommend the predicted item to the user. The target scenarios include but are not limited to e-commerce platforms where products are recommended based on the user click records in the recent period, and video streaming platforms where videos are recommended to users based on their historical watching records.

Since the records in a user-item interaction sequence are sorted in chronological order and the sequences are essentially time series, the early methods\cite{shani2005mdp,he2009web,rendle2010factorizing} model these sequences as Markov chains and predict the next actions for users based on their previous actions, thereby generating recommendations. However, these methods require strong dependency assumptions over user behaviors, and in reality, for a user, his/her next action is likely to be unrelated to the previous one but related to earlier actions. With the progress of deep learning methods, RNNs (Recurrent Neural Networks) are adopted in recommender systems due to their capabilities of modeling sequences. RNN-based methods\cite{hidasi2016parallel,hidasi2018recurrent} can capture long-term dependencies in sequences, but they are also easy to generate fake dependencies. Recently, Graph Neural Networks (GNNs), which combine the flexible expressiveness of graph data and the strong learning capability of neural networks, have emerged as a promising way to achieve recommender tasks.

On the other hand, we notice that in many recommendation scenarios, besides user-item interaction sequences, the attribute information of the items is relatively complete. Moreover, attributes of the item have been gradually used for help modeling\cite{tuan20173d,hidasi2016parallel}. However, so far, there has been still a lack of in-depth research on the modeling and mining of multiple attributes and multi-valued attributes of the item. 

To mine the potential of item attributes in learning the user preference patterns, in this paper, we treat the discrete attribute value of an item as a node on the graph. In this way, for a user-item interaction sequence, there are sequences of attributes of the item being interacted with. Next, we describe these attribute sequences with graphs, besides constructing an item graph from the user-item interaction sequence. Then, we construct a GNN model to generate next-item recommendations.

The main contributions of our work are summarized as follows.
\begin{enumerate}[topsep=1pt]
	\item We present a reasonable method to construct attribute sequences from user-item interaction sequences and attribute graphs from attribute sequences. Further, we propose a method to calculate attribute scores so as to quickly determine which attributes are valuable for modeling user preferences.
	\item We propose the sequential recommendation model Murzim. Based on gated GNNs, Murzim adopts attention mechanisms to integrate information from the node level and the sequence level, and fuses the influence of item attributes on the semantics implied in user-item interaction sequences into the recommendation results.
	\item We conduct extensive experiments on open datasets and the MX Player dataset. Murzim outperforms several methods in terms of Recall@20 and MRR@20. Moreover, we apply Murzim to the MX Player, one of India's largest streaming platforms, and the resulting business indicators such as CTR have been improved, which illustrates the effectiveness of Murzim.
\end{enumerate}

The rest of the paper is organized as follows. Section 2 introduces the related work. Section 3 gives the formulation of the problem to be solved. Section 4 describes the Murzim model in detail. Section 5 gives the experimental evaluation. Finally, the paper is concluded in Section 6.
\section{Related Work}
Conventional recommendation methods, such as item-based neighborhood methods\cite{sarwar2001item,linden2003amazon} and matrix factorization methods\cite{mnih2008probabilistic,koren2009matrix}, do not integrate with the sequential information, thus for sequential recommendation scenarios, these methods can work but perform far from the desired level.

Some existing sequence modeling methods can be adapted for the sequential recommendation. For example, in \cite{shani2005mdp}, the recommendation task is regarded as a sequence optimization problem and an MDP (Markov Decision Process) is applied to solve it. In \cite{he2009web}, a mixture variable memory Markov model is built for web query recommendation.  Further, in \cite{rendle2010factorizing}, a Markov chain model and a matrix decomposition method are combined to build the personalized probability transfer matrix for each user. However, most Markov chain based methods have to face the problem brought by the strong assumption about dependency between user behaviors.

With the rapid development of deep learning methods, RNNs demonstrate their advantages in sequential data modeling. As a result, RNNs are also adopted and then improved for the sequential recommendation\cite{hidasi2015session,hidasi2018recurrent,tan2016improved}. Besides sequential data, item features have been integrated into the RNN based models in \cite{tuan20173d,hidasi2016parallel}, where the former employs 3D convolution operations to fuse item features, and the latter employs multiple parallel RNNs with item's image features, text features, etc. as input and gives several fusion strategies. The attention mechanisms are also applied in the sequential recommendation. For example, NARM\cite{li2017neural} is a recommendation model based on an encoder-decoder structure, which designs an RNN with an attention mechanism in the encoder to capture the user's sequential behaviors and main purpose, and predicts the next item in the decoder. STAMP\cite{liu2018stamp} uses simple MLP cells and an attention network to capture the user's general and current interests to predict the next item. However, the RNN-based models have some intrinsic weaknesses since they encode the interaction sequence into a hidden vector, and using only one vector may lose information. As a remedy, some methods \cite{chen2018sequential,huang2018improving} encode user states through the memory network, which has the larger capacity. 


Recently, GNNs have received much attention, which are a kind of neural networks running on graph structure data. We note that applying the idea of CNN to the graph results in the GCN (Graph Convolution Network) methods. For example, GraphSage\cite{hamilton2017inductive} is an inductive GCN model, which aggregates node neighbor information by training a set of aggregation functions and generate the node embeddings. Applying the idea of RNN to the graph is also feasible. GGS-NN\cite{DBLP:journals/corr/LiTBZ15} is such an example. Currently, there exist several GNN models for sequential recommendations. For example, the SR-GNN model \cite{wu2019session} and its improved version \cite{xu2019graph} which borrows the self-attention structure from Transformer\cite{vaswani2017attention} and applies it to original SR-GNN.

Compared to the existing work, our work simultaneously models the user-item sequence and the corresponding attribute information in the form of directed graphs, and then build GNNs to generate the item sequence embeddings which capture the user preferences on items and attribute values.

\section{Problem Formulation}
For the sequence recommendation task, given the user-item interaction sequence set $S$, we use $V=\{v_1,v_2,\ldots,v_{|V|}\}$ to denote the set consisting of all unique items involved in all the sequences, $P=(p_{ij})_{{|V|}\times K}$ to denote the attribute matrix of the items, where $p_{ij}=f_j(v_i)\subseteq A_j$ denotes the set of values for the $j$-th attribute of the $i$-th item, $A_j=\{a^j_1,a^j_2,\ldots,a^j_m\}$ denotes the value set of the $j$-th attribute, $f_j$ is the attribute mapping function which maps the item $v_i$ into the value set $A_j, j=1,2,\ldots,K$. 

We use $s_0=[v_{s,1},v_{s,2},\ldots,v_{s,T}]$ to denote the sequence of a user's behavior over a period of time, where $v_{s,t}\in V$. At the same time, through the attribute matrix, we can get $K$ attribute sequences, that is, $s_j=[f_j(v_{s,1}),\,f_j(v_{s,2}),\,\ldots,\, \\ f_j (v_{s,T})]$. Our goal is to generate next-item prediction through the item sequence $s_0$ and $K$ attribute sequences $s_1,s_2,\ldots,s_K$, that is to predict $v_{s,T+1}$.
\section{The Murzim Model}
\begin{figure}[htb]
	\centering
	\includegraphics[width=0.72\textwidth]{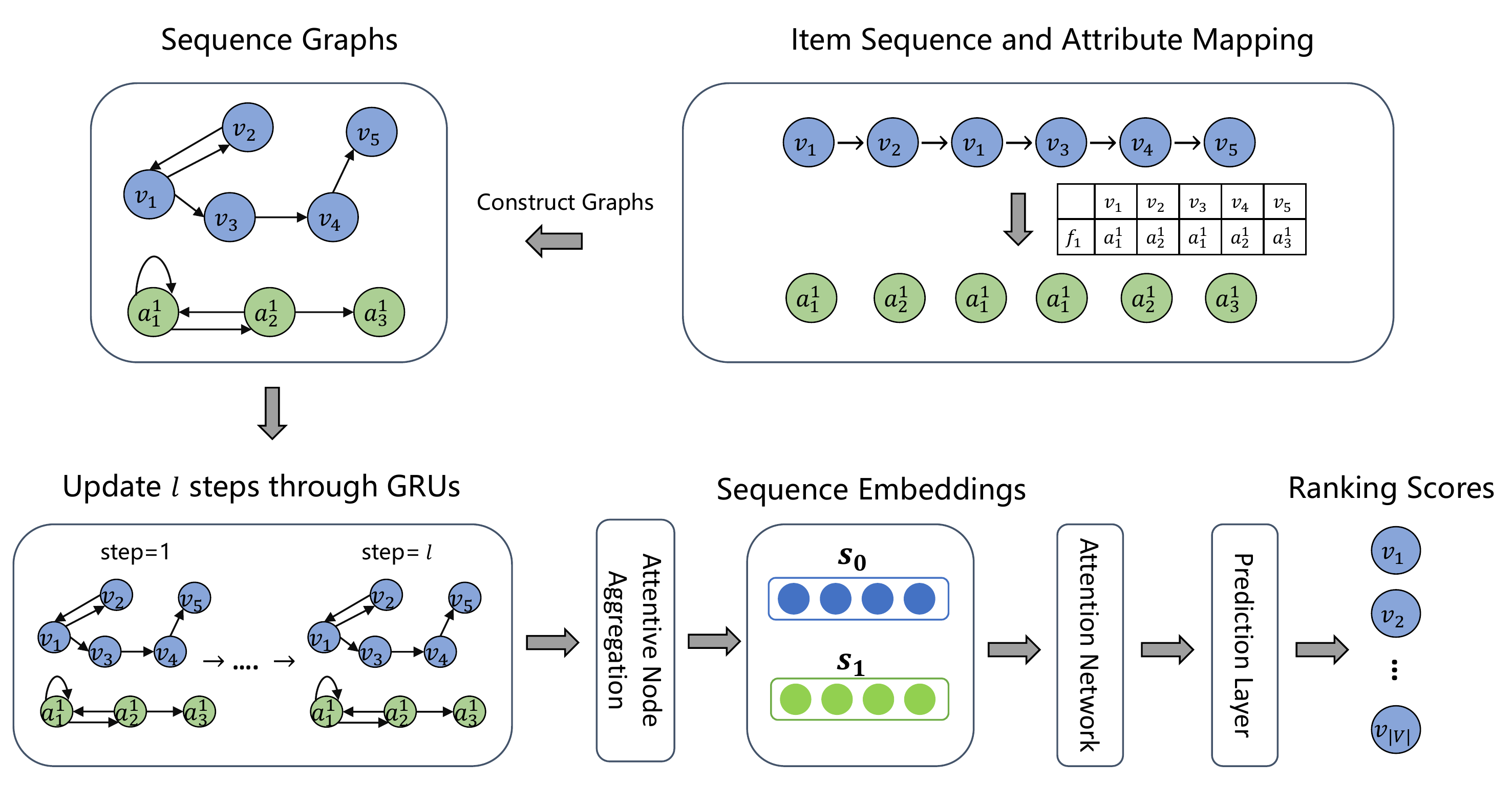}
	\caption{Structure of Murzim}
	\label{fig:fig1}
\end{figure}

In Murzim, we first construct the attribute sequences according to the item sequence and attribute mappings, and represent all the $1+K$ sequences(i.e., a item sequence and $K$ attribute sequences) as directed graphs. Then, we use gated GNN based on GRU to update the embeddings of nodes in the graphs. After obtaining the embeddings of nodes, we aggregate them to get the embeddings of the sequences. Finally, we gather all sequence embeddings through an attention network to predict the next item. The basic structure is shown in Figure \ref{fig:fig1}.
\subsection{Constructing Item and Attribute Graphs}
We use the attribute mapping functions(a total of $K$) to map the items in the user-item interaction sequence to their attribute values. In this way, we get $K$ attribute sequences. Each sequence, i.e., $s_j, j=0,1,\ldots,K$, is represented by a directed graph $G_{s_j}=(V_{s_j},E_{s_j}), V_{s_0}\subseteq V, V_{s_j}\subseteq A_j,j>0$. If two items (or attribute values) are adjacent in the sequence, we add a directed edge between the corresponding nodes in the graph, that is $\langle v_{s,t},v_{s,t+1}\rangle \in E_{s_0}$ (or $\langle f_j(v_{s,t}),f_j(v_{s,t+1})\rangle\in E_{s_j},j>0$). If $f_j$ maps an item $v_{s,t}$ to multiple attributes (for example, a movie to multiple actors), then we have a full connection between $f_j(v_{s,t})$ and $f_j(v_{s,t+1})$, where there is a directed edge between each attribute value in $f_j(v_{s,t})$ and each attribute value in $f_j(v_{s,t+1})$. 

\begin{figure}[htb]
	\centering
	\includegraphics[width=0.6\textwidth]{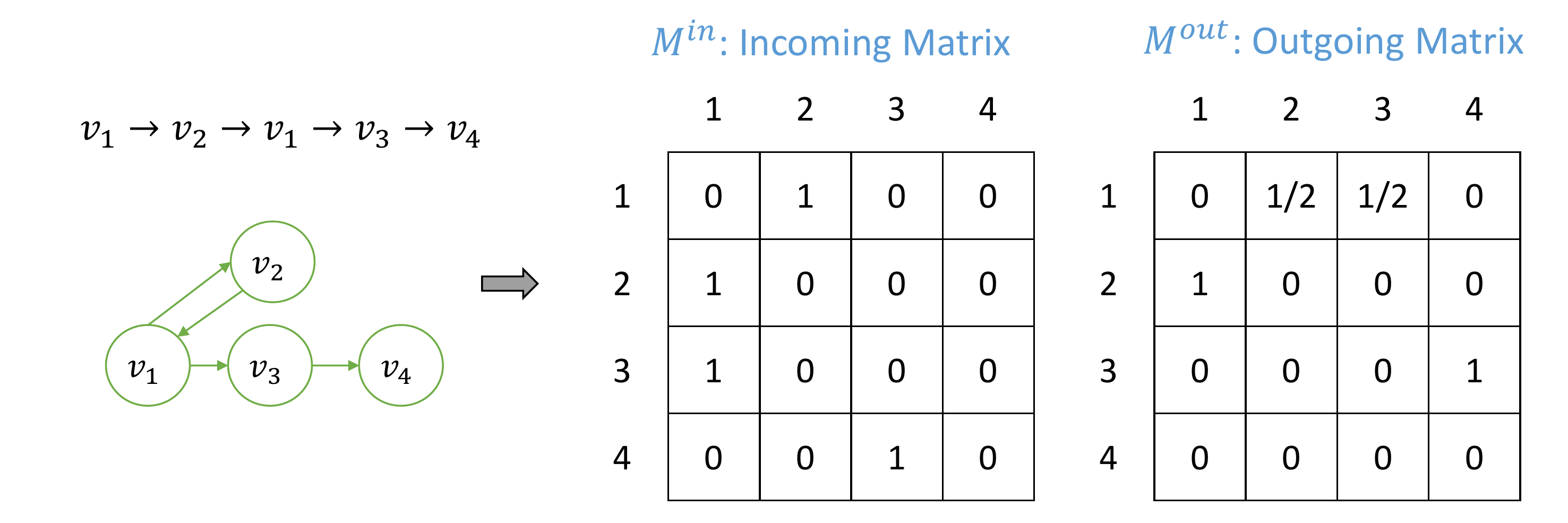}
	\caption{An attribute graph example and its corresponding matrices}
	\label{fig:fig2}

\end{figure}

We use adjacency matrices to represent the item graph and attribute graphs. Specifically, we distinguish the forward and reverse of the sequence, build incoming matrix $\bm{M^{in}}$ and outgoing matrix $\bm{M^{out}}$ respectively, and normalize each row. An example is shown in Figure \ref{fig:fig2}.

Since items often contain multiple attributes, it is necessary to assess the importance of these attributes in advance. We believe that a good attribute should reflect the user's preference on certain attribute values. To this end, we define a method to calculate the score $R(S,A_j)$ of the attribute $A_j$ on the sequence set $S$ according to the attribute map $f_j$:
\begin{equation}
R(S,A_j)=\frac{1}{|S|}\sum_{s\in S}\left(1-\frac{|\bigcup_t f_j(v_{s,t})|}{\sum_{t} |f_j(v_{s,t})|}\right)
\label{eq:attscore}
\end{equation}

In formula \eqref{eq:attscore}, $f_j(v_{s,t})$ denotes the value set of item $v_{s,t}$ in sequence $s$ on attribute $j$, $|\bigcup_t f_j(v_{s,t})|$ denotes the number of values of all items in sequence $s$ on attribute $j$, and $\sum_{t} |f_j(v_{s,t})|$ denotes the sum of the number of values of each item in sequence $s$ on attribute $j$.  The greater the number of attribute values in a sequence, that is, the larger $\sum_{t} |f_j(v_{s,t})|$, and the more concentrated these attribute values, that is, the smaller $|\bigcup_t f_j(v_{s,t})|$, the higher the attribute score, which means the user's interest is focused on only a few attribute values. When we apply Murzim in a specific scenario, we first calculate the scores and then select the attributes with high scores for modeling.

\subsection{Generating Node Embeddings}
\label{sec:gne}
We get the initial item embeddings through the embedding look-up operations from a trainable matrix with dimension $d \times |V|$. The $d$-dimensional vector $\bm{v_i}$ is used to represent the embedding of the $i$-th item. As for attribute, we treat it as a partition of items, that is, different attribute values divide items into multiple intersecting or disjoint sets. Therefore, we use item embeddings to generate the initial embeddings of attribute values.

Let $s_0=[v_{s,1},v_{s,2},\ldots,v_{s,T}]$ be the user-item interaction sequence, $s_j=[f_j(v_{s,1}),f_j(v_{s,2}),\ldots,f_j (v_{s,T})]$ be the sequence corresponding to attribute $j$, and $A_{s,j}=\cup_{i=1}^Tf_j(v_{s,i})$ be the value set of attribute $j$ on the item sequence $s_0$, $A_{s,j}\subseteq A_j$. For any $a^j\in A_{s,j}$, the calculation of its embedding $\bm{a^j}$ is:
\begin{equation}
\bm{a^j}=\frac{1}{|V_{s_0}|}\sum_{i=1}^{|V_{s_0}|}I(a^j\in f_j(v_{i}))\bm{W_jv_{i}}
\label{eq:att_emb}
\end{equation}

In formula \eqref{eq:att_emb}, $V_{s_0}$ is the set of items contained in the user-item interaction sequence $s_0$,  $\bm{W_j}\in \mathbb{R}^{d\times d}$ is a model parameter, $I$ is an indicator function that outputs 1 when its input is true, otherwise outputs 0.  

After obtaining the initial embeddings of items and attribute values, we update them through graphs. Let $\bm{e_i}$ represent the embedding of node $i$ on the graph (item graph or attribute graph), and then we propagate the information between nodes according to the matrices $\bm{M^{in}}$ and $\bm{M^{out}}$, as shown below.
\begin{equation}
\bm{m_i}=\text{concat}(M^{in}_{i*}[\bm{e_1}, \ldots, \bm{e_n}]^T, M^{out}_{i*}[\bm{e_1}, \ldots, \bm{e_n}]^T)
\label{eq:msg_prop}
\end{equation}

In formula \eqref{eq:msg_prop}, $[\bm{e_1}, \ldots, \bm{e_n}]$ is a matrix of size $d\times n$ formed by the embeddings of all $n$ nodes in the graph, $M^{in}_{i*},M^{out}_{i*}\in \mathbb{R}^{1\times n}$ denote the $i$-th row of the corresponding matrix. In the subsequent step, $\bm{m_i}\in \mathbb{R}^{2d\times 1}$ is used as the input of the GRU to updated the embedding of node $i$:
\begin{equation}
	\bm{e_i} = \text{GRU}(\bm{e_i}, \bm{m_i})
\end{equation}

The above process is iterated multiple times, so that each node can obtain information from nodes farther away.

\subsection{Generating Sequence Embedding}
The generation of sequence embeddings goes through two steps. We first aggregate the embeddings of the nodes in different graphs to get the embeddings of $1+K$ sequences. There are many ways to aggregate, here we choose the method similar to \cite{wu2019session}:  calculate the attention coefficient of each node embedding with the last item embedding(or attribute value embedding) in the sequence, and sum all node embeddings according to the coefficients:
\begin{gather}
	\alpha_i=\bm{q}^T\sigma(\bm{W_1}\bm{e_T}+\bm{W_2}\bm{e_i}+\bm{c}) \label{eq:sq_emb1} \\
	\bm{\tau}=\sum_{i=1}^{n}\alpha_i\bm{e_i} \label{eq:sq_emb2}\\
	\bm{s_j}=\bm{W_3}\text{concat}(\bm{\tau};\bm{e_T})  \label{eq:sq_emb3}
\end{gather}

In formulas \eqref{eq:sq_emb1}-\eqref{eq:sq_emb3}, $\bm{W_1},\bm{W_2}\in\mathbb{R}^{d\times d},\bm{W_3}\in\mathbb{R}^{d\times 2d},\bm{q},\bm{c}\in\mathbb{R}^d$ are model parameters, $\bm{e_T}$ is the embedding of the last item(or attribute value) in the sequence, the output $\bm{s_j}$ is the embedding of the sequence($j=0$ for item sequence, $j=1,\ldots, K$ for attribute sequences).

Then we use an attention network to aggregate all sequence embeddings, and finally get the embedding $\bm{z}$ containing both the user's item preference and attribute preference:
\begin{gather}
	\alpha_j=\sigma(\frac{\bm{s_0}\bm{W_q}\bm{s_j}^T}{\sqrt{d}}), j=0,1,2,\ldots K
	\label{att_coef} \\
	\bm{z}=\bm{s_0}+\sum_{j=0}^{K}\alpha_j\bm{s_j}
	\label{att_sum}
\end{gather}

In formula \eqref{att_coef}, the item sequence embedding $\bm{s_0}$ is used as the query in the attention mechanism, $\bm{W_q}\in \mathbb{R}^{d\times d}$ is a model parameter. We perform weighted summation in formula \eqref{att_sum} to get $\bm{z}$ that integrates all sequence information.

\subsection{Generating Prediction and Model Loss}
We use cosine similarity to score all items according to $\bm{z}$:
\begin{equation}
	\hat{y}_i=softmax\left(\gamma\frac{\bm{z}^T\bm{v_i}}{||\bm{z}||_2||\bm{v_i}||_2}\right)
	\label{eq:item_score}
\end{equation}

In formula \eqref{eq:item_score}, $\gamma$  is a trainable factor, and item embedding $\bm{v_i}$ is obtained in Section \ref{sec:gne}. We adopt the cross entropy loss for predicting the next item:
\begin{equation}
L=-\sum_{i=1}^{|V|}y_i\log\hat{y}_i +\lambda||\theta||^2
\label{eq:loss}
\end{equation}

In formula \eqref{eq:loss}, $y\in\mathbb{R}^{|V|}$ is the one-hot encoding vector corresponding to the ground truth item. $\theta$ is the set of all trainable parameters of the model. We train the model by optimizing $L$ through the gradient descent method.
\section{Experiments and Analyses}
To evaluate the performance of Murzim, we conduct performance comparison experiments on different datasets, comparing Murzim with several existing models. Next, we deploy Murzim to the MX player and observe its performance in the actual production environment.
\subsection{Experimental Setup}
\noindent\textbf{Datasets:} We adopt the Yoochoose dataset, the Diginetica dataset, and our own MX Player dataset.

The Yoochoose dataset is from RecSys Challenge 2015\footnote{http://2015.recsyschallenge.com/challege.html}. It provides the user click sequence data of an e-commerce website, and contains the category information for each item which is treated as an attribute in our model. In particular, we fetch recent 1/64 and 1/4 sequences of the total Yoochoose dataset to form two datasets. The Diginetica dataset is from CIKM Cup 2016\footnote{http://cikm2016.cs.iupui.edu/cikm-cup}, which provides more item attribute information, including: category, priceLog2 (log-transformed product price) and item name token (comma separated hashed product name tokens). We only use its transactional data. 

As for the pre-processing of the two datasets, we keep the same as \cite{wu2019session}. Further, the training and testing sets are divided in the same way as \cite{wu2019session}: data in the last day in the Yoochoose dataset are used as the testing set, and data in the last week of the Diginetica dataset are used as the testing set.

Besides, we construct a dataset named MXPlayer\_1W\_1M from the MX Player log. We extract the interaction sequences on the movies from 2020-01-20 to 2020-01-26, and filter out items whose number of occurrences is less than 3 and sequences whose length is 1. We use the data in the first six days for training and the last day for testing. Here, the items, i.e., movies, contain rich attribute information. We select seven attributes of items: genre (G), publisher (P), country(C), language(L), release year(R), director(D), and actor(A).

The details of the three datasets are shown in Table \ref{tab:yoo_digi_dataset}.
\begin{table}[htb]
	\centering
	\caption{Details of the three datasets.}
	\label{tab:yoo_digi_dataset}
	\scriptsize
	\begin{tabular}{lccccccccccc}
		\toprule
		Dataset     & \#Item                 & \#Train                  & \#Test                 & Avg.Len               & \multicolumn{7}{c}{\#Attribute Values}                                                     \\
		\midrule
		\multirow{2}{*}{Yoochoose 1/64} & \multirow{2}{*}{17745} & \multirow{2}{*}{369859}  & \multirow{2}{*}{55898} & \multirow{2}{*}{6.16} & \multicolumn{7}{c}{category}                                   \\ 
		&                        &                          &                        &                       & \multicolumn{7}{c}{282}                                                                             \\  \cmidrule{1-12}
		\multirow{2}{*}{Yoochoose 1/4}  & \multirow{2}{*}{30470} & \multirow{2}{*}{5917745} & \multirow{2}{*}{55898} & \multirow{2}{*}{5.71} & \multicolumn{7}{c}{category}                          \\  
		&                        &                          &                        &                       & \multicolumn{7}{c}{322}                                                                             \\ \cmidrule{1-12}
		\multirow{2}{*}{Diginetica}     & \multirow{2}{*}{43097} & \multirow{2}{*}{719470}  & \multirow{2}{*}{60858} & \multirow{2}{*}{4.85} & \multicolumn{2}{c}{category} & \multicolumn{3}{c}{priceLog2} & \multicolumn{2}{c}{name token} \\  
		&                        &                          &                        &                       & \multicolumn{2}{c}{1217}       & \multicolumn{3}{c}{13}          & \multicolumn{2}{c}{164774}       \\  \cmidrule{1-12}
		\multirow{2}{*}{MXPlayer\_1W\_1M}      & \multirow{2}{*}{11386} & \multirow{2}{*}{1658463} & \multirow{2}{*}{69804} & \multirow{2}{*}{6.38} & G            & P           & C       & L      & R      & D            & A             \\
		&                        &                          &                        &                       & 34             & 130           & 55        & 40       & 106      & 6806           & 16248        \\  
		\bottomrule  
	\end{tabular}
\end{table}

According to the attribute score calculation method defined previously, we obtain attribute scores on the training sets of different datasets, as shown in Table \ref{tab:yoo_digi_as}. For the Diginetica dataset, the category attribute has the highest score, which indicates that most users seem to only pay attention to a few categories of items during browsing. For the MXPlayer\_1W\_1M dataset, the language attribute has the highest score, indicating that the values of language might play more important roles in the sequence.

\begin{table}[tb]
	\centering
	\caption{Attribute scores on the three datasets}
	\label{tab:yoo_digi_as}
	\scriptsize
	\begin{tabular}{lccccccc}
		\toprule
		Dataset    & \multicolumn{7}{c}{Attribute Scores}                                                                    \\
		\midrule
		\multirow{2}{*}{Yoochoose 1/64} & \multicolumn{7}{c}{category}                                                                  \\
		& \multicolumn{7}{c}{0.5962}                                                                    \\ \cmidrule{1-8}
		\multirow{2}{*}{Yoochoose 1/4}  & \multicolumn{7}{c}{category}                                                                  \\
		& \multicolumn{7}{c}{0.5974}                                                                    \\  \cmidrule{1-8}
		\multirow{2}{*}{Diginetica}     & \multicolumn{2}{c}{category} & \multicolumn{3}{c}{priceLog2} & \multicolumn{2}{c}{name token} \\
		& \multicolumn{2}{c}{0.6450}   & \multicolumn{3}{c}{0.4971}    & \multicolumn{2}{c}{0.2498}     \\  \cmidrule{1-8}
		\multirow{2}{*}{MXPlayer\_1W\_1M}      & G             & P            & C        & L        & R       & D              & A             \\
		& 0.4218        & 0.4155       & 0.5385   & 0.6271   & 0.2648  & 0.1719         & 0.1833       \\
		\bottomrule
	\end{tabular}
\end{table}

\noindent \textbf{Metrics:} We adopt Recall@20 and MRR(Mean Reciprocal Rank)@20 to evaluate the recommendations. Recall@20 is the proportion of ground-truth items among the top-20 recommended items, and here is equivalent to Hit@20. MRR@20 is the average of the inverse of the ranking of ground truth in the recommendation results. If the ground truth does not appear in the top-20 of the recommendations, then MRR@20 is 0.

\noindent \textbf{Implementation Details:} In Murzim, the embedding dimension $d$ is set to 64 and 128 on the public datasets and MXPlayer\_1W\_1M dataset, respectively, the L2 penalty and batch size are set to 1e-5 and 512, respectively. We implement Murzim with Tensorflow, using the Adam optimizer where the initial learning rate is set to 0.004 and decays by 0.1 after every 2 epochs.

\subsection{Performance Comparison}
First, we conduct a comparative experiment, comparing our model with the following six models:
\begin{enumerate}[topsep=0pt]
	\item POP/S-POP: They recommend the top-N popular items in the entire training set or the current sequence. In S-POP, if the number of recommended items is insufficient, we use top-N popular items in the entire training set for completion.
	\item Item-KNN\cite{sarwar2001item}: It recommends the top-N items that are most similar to the items in the sequence. The similarity between item $i$ and item $j$ is calculated based on the number of co-occurrences in the sequence.
	\item GRU4Rec\cite{hidasi2015session}: It models sequences with RNN to predict the next item.
	\item NARM\cite{li2017neural}: It adds an attention mechanism to RNN to capture user's sequence behavior and main interaction purpose in the current sequence.
	\item STAMP\cite{liu2018stamp}: It uses a new attention mechanism to capture general interest and short-term attention of users.
	\item SR-GNN\cite{wu2019session}: It uses a GNN to model sequences, while using an attention mechanism to fuse users' long-term and short-term interests in sequences.
\end{enumerate}
\begin{table}[tb]
	\centering
	\caption{Performance comparison}
	\label{tab:pub_datasets_ret}
	\scriptsize
	\begin{tabular}{lcccccccc}
		\toprule
		\multirow{2}{*}{} & \multicolumn{2}{c}{Yoochoose 1/64} & \multicolumn{2}{c}{Yoochoose 1/4} & \multicolumn{2}{c}{Diginetica} & \multicolumn{2}{c}{MXPlayer\_1W\_1M} \\ 
		& Recall@20            & MRR@20               & Recall@20            & MRR@20              & Recall@20          & MRR@20   & Recall@20          & MRR@20          \\ 
		\midrule
		{POP}            & 6.71                 & 1.65                 & 1.33                 & 0.30                & 0.89               & 0.20    & 14.78 & 3.95         \\
		{S-POP}          & 30.44                & 18.35                & 27.08                & 17.75               & 21.06              & 13.68    & 29.14 & 13.67            \\
		{Item-KNN}       & 51.60                & 21.81                & 52.31                & 21.70               & 35.75              & 11.57     & 50.03 & 19.33         \\
		{GRU4Rec}        & 60.64                & 22.89                & 59.53                & 22.60               & 29.45              & 8.33    & 53.54 & 20.71             \\
		{NARM}           & 68.32                & 28.63                & 69.73                & 29.23               & 49.70              & 16.17       & 53.97 & 20.43         \\
		{STAMP}          & 68.74                & 29.67                & 70.44                & 30.00               & 45.64              & 14.32     & 53.80 & 20.56           \\
		{SR-GNN}         & 70.57                & 30.94                & 71.36                & 31.89               & 50.73              & 17.59       & 55.05  & 21.52         \\
		{Murzim}            & \textbf{71.52}          & \textbf{31.65}         & \textbf{72.19}          & \textbf{32.04}         & \textbf{54.74}              & \textbf{19.40}      &  \textbf{55.51}  &  \textbf{21.82}          \\
		\bottomrule
	\end{tabular}
\end{table}

The experimental results are shown in Table \ref{tab:pub_datasets_ret}.  We note that SR-GNN works best among the comparison methods, which shows the effectiveness of graph-based representation in the sequential recommendation. However, our model (i.e., the Murzim version with the category attribute for public datasets and language attribute for MX Player dataset) outperforms SR-GNN. The performance gain over SR-GNN should originate from the fact that Murzim mines user preferences implicit in the attributes of items that users interact with.

Next, we observe how and to what degree Murzim exploits the attribute effects. Table \ref{tab:dig_att_comb} shows the performance of different versions of Murzim on the Diginetica dataset, where the different versions adopt different attributes or attribute combinations. The results in the first three lines are consistent with the attribute scores we calculate above: the higher the score is, the greater the performance improvement is. The results in the subsequent lines show the effect of adding multiple attributes. We find that except for the combination of pricelog2+name token, the effects of other combinations are lower than just adding one attribute. It indicates that attributes might influence each other and the performance is not always improved with the increasing of attribute information. 

\begin{table}[tb]
	\centering
	\caption {Effects of different attribute combinations on the Diginetica dataset}
	\label{tab:dig_att_comb}
	\scriptsize
	\begin{tabular}{lcc}
		\toprule
		{Attribute Combination} & {Recall@20} & {MRR@20} \\ \midrule
		category                          & \textbf{54.74}              & \textbf{19.40}           \\
		pricelog2                         & 54.29              & 19.19           \\
		name token                       & 54.27               & 19.20           \\
		category + pricelog2              & 54.60     & 19.36           \\
		pricelog2 + name token            & 54.60              & 19.40           \\
		category + name token             & 54.32              & 19.24           \\
		category + pricelog2 + name token & 54.66              & 19.39  \\ 
		\bottomrule
	\end{tabular}
\end{table}

Table \ref{tab:mx_ret} shows the performance of different versions of Murzim on the MXPlayer\_1W\_1M dataset. The experimental results in the first seven columns are generally consistent with the scores of attributes of the MXPlayer\_1W\_1M dataset in Table \ref{tab:yoo_digi_as}. For example, the language attribute score is the highest, and the corresponding version of Murzim works best in comparison with the other versions with a single attribute. This shows that our attribute score calculation method can measure the importance of attributes to a certain extent. Then, we select any two attributes from three attributes with the highest scores: language, country, and genre, and form the corresponding version of Murzim to conduct the experiment. The results are listed in the last three columns in Table \ref{tab:mx_ret}. We find that the combination of any two attributes does not get further improvement on MRR@20. Meanwhile, although the combination of language and genre achieves the best result on Recall@20, the results of the other groups are close to that of adding a single attribute. This indicates again that attributes are useful in improving the recommendation but it is not necessarily the case that more attributes lead to better recommendations.
\begin{table}[tb]
	\centering
	\caption {Effects of different attribute combinations on the MXPlayer\_1W\_1M dataset}
	\label{tab:mx_ret}
	\scriptsize
	\begin{tabular}{lcccccccccc}
		\toprule
		& G & P & C & L & R & D & A & L+C & L+G & C+G \\
		\midrule
		Recall@20 & 55.44 & 55.45 & 55.43 & 55.51 & 55.27  & 55.31 & 55.16 & 55.50  & 55.47  & \textbf{55.54} \\
		MRR@20  & 21.71 & 21.74 & 21.76 & \textbf{21.82} & 21.70 & 21.73 & 21.61 & 21.77 & 21.80 & 21.75 \\
		\bottomrule
	\end{tabular}
\end{table}

\subsection{Online Test}
\begin{figure}[htb]
	\centering
	\includegraphics[width=0.5\textwidth]{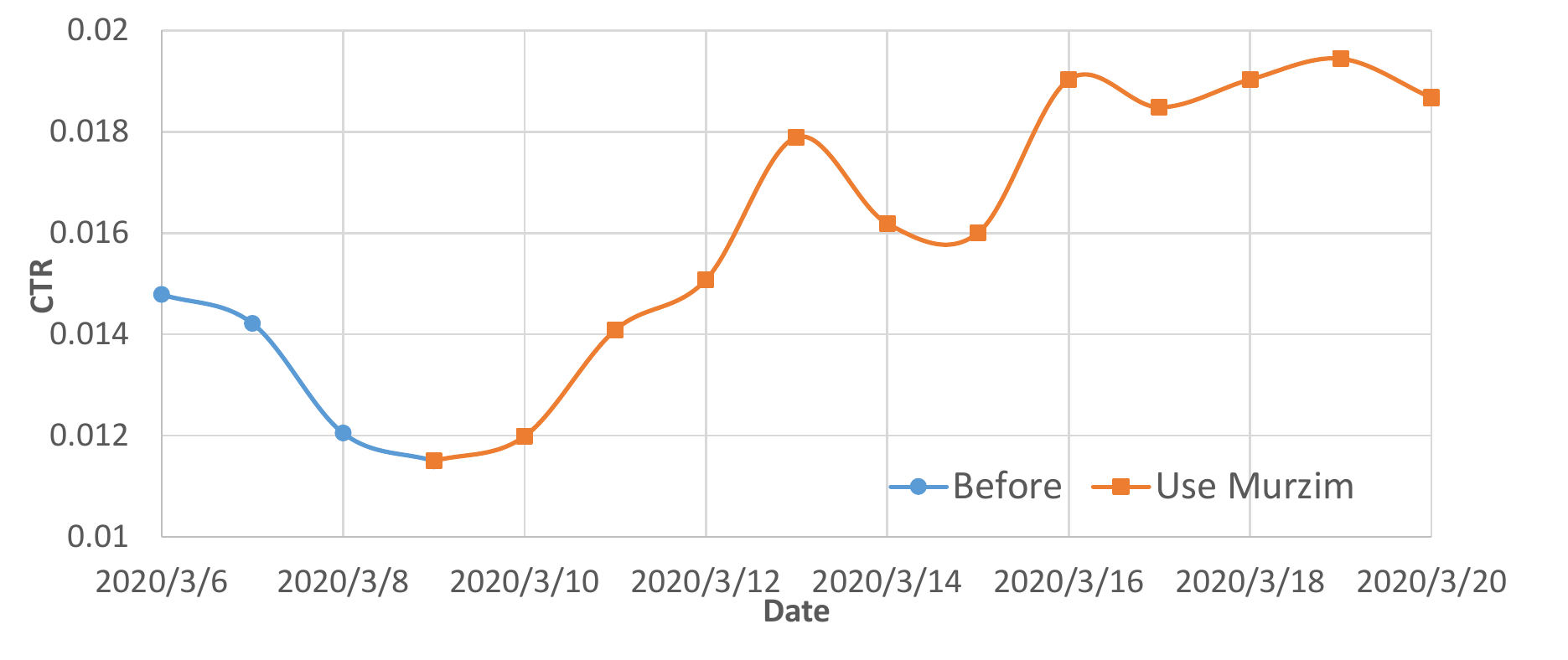}
	\caption{ CTRs before and after using Murzim}
	\label{fig:fig3}
\end{figure}

We have deployed Murzim to the online production environment of the MX Player and generate recommendation for users based on their viewing sequences. In the online version of Murzim, we choose two attributes, i.e., language and genre. We compare the click-through rates (CTRs) on a group of users before and after using Murzim, as shown in Figure \ref{fig:fig3}. It can be seen that after Murzim is deployed, the CTR gradually increases. Compared with the previous values, it increases by about 60\% on the average. This shows that Murzim generates better recommendations than before. 

\section{Conclusion}
In this paper, we propose a GNN model Murzim for sequential recommendation. Murzim describes the user-item interaction sequences and attribute sequences by directed graphs, and then gathers information through node aggregation and the attention network, which not only inherits the advantages of GNNs but also excavates user preferences through attributes, thus improving the recommendation performance. Currently, Murzim is running in the online production environment of the MX player and mainly serving the people in India.

\section*{Acknowledgement}
This work was supported by the National Natural Science Foundation of China under Grant No. 62072450 and the 2019 joint project with MX Media.
%
%
%
%
\bibliographystyle{splncs04}
\bibliography{ref.bib}
\end{document}